\documentstyle[12pt]{article}

\def\tbar{{\bar t}}
\def\bPsi{{\bar\Psi}}

\def\tr{{\rm \;tr\;}}
\def\det{{\rm det\;}}
\def\log{{\rm log\;}}
\def\hM{{\hat M}}
\def\hm{{\hat m}}

\begin{document}
\hsize37truepc\vsize61truepc
\hoffset=-.5truein\voffset=-0.8truein
\setlength{\baselineskip}{17pt plus 1pt minus 1pt}
\setlength{\textheight}{22cm}

\begin{titlepage}
\begin{flushright}
hep-th/9511127\\
TIFR/TH/95-55\\
November 1995\\
\end{flushright}
\vspace{2cm}
\begin{center}
{\large\bf Matrix Models, Quantum Penner Action\\
\vspace{0.3cm}
and Two-Dimensional String Theory\footnote{Talk given by the second
author at the Nato Advanced Summer Institute on
{``\it Low Dimensional Applications of Quantum Field Theory''},
Cargese, July 11-29 1995.
}\\}
\vspace{0.8cm}
{\bf Camillo Imbimbo}\\
\vspace{0.3cm}
{\it INFN, Sezione di Genova}\\
{\it I-16146 Genova, Italy}\\
\vspace{0.8cm}
{\bf Sunil Mukhi}\\
\vspace{0.3cm}
{\it Tata Institute of Fundamental Research}\\
{\it Bombay 400 005, India}
\end{center}
\vspace{2cm}
\begin{center}
ABSTRACT
\end{center}
{\parindent=0pt A very elementary model of a single positive hermitian
random matrix coupled to an external matrix is defined and
studied. Expanding the exact effective action around its classical
solution leads to the ``quantum Penner action'', from which a rich
structure of correlation functions is obtained. These are shown to be
equal to the all-orders perturbative expansion of tachyon amplitudes
in the two-dimensional string at self-dual radius.}
\end{titlepage}

\section{Introduction}

Random-matrix models have been intensively studied in the last few
years\cite{ginsmoore}. Although the bulk of the attention has been
paid to models describing dynamically triangulated random
surfaces\cite{kazakov}, and their double-scaling limits\cite{double},
it has turned out that many of the results obtained can be
equivalently deduced from a far simpler class of matrix models, which
moreover require no double-scaling limit. These are sometimes called
``topological matrix models'' because the random-matrix integral can
be thought of as a generating function for certain topological
invariants.

The first such model was the one constructed by
Penner\cite{penner}\cite{distvaf} to study the
virtual Euler characteristic of the moduli space of Riemann
surfaces. Subsequently, Kontsevich\cite{konts} obtained a model which
generates intersection numbers on this moduli space (see also
Ref.\cite{morekonts}), and generalizations of the Kontsevich
model\cite{genkonts} were soon found which describe more complicated
topological problems associated to vector bundles over moduli space.

The simplest of this class of models will be the subject of this
article. We will show that a simple (almost trivial) model of
hermitian {\em positive} random matrices coupled to an external matrix
gives rise to a fascinating theory\cite{im} which can eventually be
seen to generalize both the Penner and Kontsevich models. Moreover,
our theory will be shown to satisfy $W_\infty$ identities which
uniquely fix its perturbative expansion to be that of noncritical
$c=1$ string theory at self-dual radius\cite{dmp}\cite{gim}.

The two-dimensional string at this radius has been recently
argued\cite{gv} to determine universal properties of type II
superstrings compactified on Calabi-Yau manifolds with conifold
singularities. This suggests the exciting possibility
that the matrix model described here could have a direct bearing on
physical properties of type II superstrings. We will make some
speculations in this and other directions below.

\section{A Topological Matrix Model}

Consider a single $N\times N$ hermitian matrix $M$ whose eigenvalues
are constrained to be positive-semidefinite. We will choose a {\em
linear} action, describing the coupling of $M$ to a fixed external
matrix $A$. Remarkably, this almost trivial choice leads us to a
matrix model which describes a well-known string theory, as we will
see below.

The random-matrix integral is
\begin{equation}
Z(A) = \int dM~ e^{-\nu\tr MA}
\label{random}
\end{equation}
where $\nu$ is a coupling constant. The eigenvalue part of the
integral is taken over the range $[0,\infty)$. Thus this integral is
convergent.

The redefinition $M\rightarrow MA^{-1}$ transfers the $A$-dependence
to a prefactor:
\begin{equation}
Z(A) = (\det A)^{-N}\int dM~ e^{-\nu\tr M}
\label{prefactor}
\end{equation}

We would like to think of this as a toy version of a quantum field
theory. Treating $A$ as a fixed source (a background), we add an extra
source term and compute the Legendre transform of the free energy to
obtain the effective action. Thus we start now with
\begin{equation}
Z_A(J) = \int dM~ e^{-\nu\tr MA - \tr JM} =
\big(\det(A+{J\over\nu})\big)^{-N} \int dM~ e^{-\nu\tr M}
\label{source}
\end{equation}
The free energy as a function of the background $A$ and the source $J$
is minus the log of this integral, which is (dropping additive
constants):
\begin{equation}
F_A(J) = N\tr\log\big(A+{J\over\nu}\big)
\label{free}
\end{equation}

To find the effective action, define the ``classical field''
$\hM$ by
\begin{equation}
\hM \equiv {\partial F_A(J)\over\partial J} =
{N\over\nu}\big(A+{J\over\nu}\big)^{-1}
\label{qfield}
\end{equation}
Then make the Legendre transformation
\begin{equation}
\Gamma(\hM) = F_A(J) - \tr \hM J
\label{legendre}
\end{equation}
and eliminate $J$ from Eq.(\ref{qfield}) to get (again dropping an
additive constant)
\begin{equation}
\Gamma(\hM) = \nu\tr\hM A - N\tr\log\hM
\label{effac}
\end{equation}

It is perhaps surprising that in such a trivial theory, the
effective action is not identically equal to the classical one! The
difference is an additive logarithmic term, which is generated
dynamically --- essentially by the boundary of the integration region
at 0. The coefficient $N$ in front of the log term is undesirable
since we ultimately intend to take the limit of large $N$ with the
coupling $\mu$ fixed. It can be easily removed by including in the
``bare'' action a term of the same logarithmic type, with
coefficient $(\nu-N)$.

This finally leads us to consider the model with matrix integral
\begin{equation}
Z(A) = \int dM~ e^{-\nu\tr MA +(\nu-N)\tr\log M}
\label{finalmod}
\end{equation}
whose effective action is
\begin{equation}
\Gamma(\hM) = \nu(\tr\hM A - \tr\log\hM)
\label{}
\end{equation}
Continuing to treat this as a toy field theory, we will show right
away that this action gives rise to a rich set of
``amplitudes''. Later we will see that these turn out to be precisely
the {\em tachyon scattering amplitudes of $c=1$ string theory} at the
self-dual radius, with ${1\over\nu^2}$ as the genus-expansion
parameter!

To evaluate amplitudes, we need to first solve the equations of motion
coming from the effective action $\Gamma(\hM)$:
\begin{equation}
0 = {\partial\Gamma\over\partial\hM} = A - \hM^{-1}
\label{eqofm}
\end{equation}
and then shift the ``field'' $\hM$ about the classical solution
$\hM=A^{-1}$ by $\hM=A^{-1}+\hm$. Then $\hm$ is the ``quantum field'',
and its amplitudes are determined by the expansion of the effective
action $\Gamma(\hM)$ about the classical solution, which leads to the
``quantum Penner action'':
\begin{equation}
\Gamma(\hm) = \nu\tr\log A + \nu \sum_{k=2}^\infty {(-1)^k\over k}\tr
(A\hm)^k
\label{qpenn}
\end{equation}
For the special case $A=1$ this coincides with the classical action of
the Penner model, whose partition function computes the virtual Euler
characteristic of the moduli space of Riemann surfaces. Here we see
that a general background $A$ gives rise to a full-fledged theory of
amplitudes. To compute these, we read off the 1PI vertices, for
example the two- and three-point vertices are
\begin{eqnarray}
\Gamma^{(2)}_{i_1j_1;i_2j_2} &=& \nu\,A_{i_2j_1}A_{i_1j_2}\\
\Gamma^{(3)}_{i_1j_1;i_2j_2;i_3j_3}&=&
-\nu\,\left[ A_{i_3j_1}A_{i_1j_2}A_{i_2j_3} +
A_{i_2j_1}A_{i_3j_2}A_{i_1j_3}\right]
\label{vertices}
\end{eqnarray}
while the propagator is the inverse of the two-point vertex:
\begin{equation}
G^{(2)}_{i_1j_1;i_2j_2} = \langle \hm_{i_1j_1} \hm_{i_2j_2}\rangle =
{1\over \nu} A^{-1}_{i_1j_2} A^{-1}_{i_2j_1}
\label{prop}
\end{equation}

Now we can compute $n$-point functions in terms of the background
$A$. At this point we take the limit of large $N$, and parametrize the
background in terms of an infinite number of independent parameters
$t_n$ via the Kontsevich-Miwa transform
\begin{equation}
t_n = {1\over\nu}\tr{A^{-n}\over n}
\label{kmtrans}
\end{equation}
Then the two- and three-point functions are easily computed to be
\begin{eqnarray}
\langle \tr \hM^2\rangle &=& \nu(2 t_2 + (t_1)^2)\\
\langle \tr \hM^3\rangle &=& \nu(3 t_3 + 6 t_1 t_2 + (t_1)^3 +
{1\over\nu^2}3 t_3)
\label{twothree}
\end{eqnarray}
Higher point functions follow using the higher vertices calculated
from the action, by computing all connected and disconnected tree
diagrams.

The structure of the amplitudes already begins to look
interesting. Because of obvious homogeneity properties of the quantum
Penner action, the $n$-point function will be a quasi-homogeneous
function of the $t_i$ where each term has $\sum i =n$. Moreover, apart
from an overall factor of $\nu$, the amplitudes have the form of a
power series expansion in $1\over\nu^2$ which {\em terminates} at
order $\left({1\over\nu^2}\right)^{[{n-1\over 2}]}$ where $[~]$ denotes
the integer part. In the next section we show that the amplitudes that
we obtain in this way represent the complete perturbative solution of
$c=1$ string theory at the self-dual radius, with the operators $tr
\hM^n$ representing the tachyons of momentum $-n$ in that theory, and
the parameters $t_n$ representing the couplings to tachyons of
positive momentum.

\section{Relation to $c=1$ string}

The $c=1$ string is a background of bosonic string theory where the
spacetime is two-dimensional. One of these dimensions may be compact,
and we focus on the case where this compact direction has radius unity
in suitable units, the self-dual value under $T$-duality. The
perturbative solution of this theory takes the form of a generating
function for ``tachyon'' scattering amplitudes, satisfying a set of
recursive $W_\infty$ ward identities which completely determine
it (at the self-dual radius, these have been obtained from matrix
models in Ref.\cite{dmp} and from the topological Landau-Ginzburg
model in Ref.\cite{gim}). Denote the partition function of this theory
$Z_{W_\infty}(t,\tbar)$ where $t_n, \tbar_n$ ($n=1,2,\ldots$) are
respectively the couplings to tachyons of positive and negative
momentum $|n|$. The $W_\infty$ identities can be written
\begin{equation}
{1\over \mu^2}{\partial Z_{W_\infty}\over \partial
\tbar_n}(t,\tbar) = {1\over (n+1)(i\mu)^{n+1}}\oint dz\, W_{n+1}(z,t,
{\partial\over\partial t})~ Z_{W_\infty}
\label{winf}
\end{equation}
Here, $\mu$ is the cosmological constant of the theory, and the
$W$-generators are differential operators defined in terms of certain
free-fermion operators by
\begin{equation}
W_{n+1} =~ :\! \bPsi\,\partial_z^{n+1}\,\Psi\! :
\label{wgen}
\end{equation}
The free-fermion operators in turn are defined through bosonization,
\begin{eqnarray}
&&\Psi(z,t,{\partial\over\partial t}) =
e^{i\mu\phi(z,t,{\partial\over\partial t})}\\
&&\bPsi(z,t,{\partial\over\partial t}) =
e^{-i\mu\phi(z,t,{\partial\over\partial t})}\\
&&\partial \phi(z,t,{\partial\over\partial t}) =
{1\over z} + \sum_{n>0} n t_n z^{n-1} -
{1\over \mu^2} \sum_{n>0} {\partial\over \partial t_n} z^{-n-1}
\label{boson}
\end{eqnarray}

To make contact with the matrix model of the previous section, we
replace the couplings $t_n$ by an external hermitian matrix $A$
defined through Eq.(\ref{kmtrans}) above. The limit
$N\rightarrow\infty$ on this matrix is implicit, since this is
required for the parameters $t_n$ to be all independent. A lengthy but
straightforward computation\cite{im} shows that the $W_\infty$
constraint can be rewritten
\begin{equation}
{1\over i\mu}{\partial Z_{W_\infty}\over \partial
\tbar_n}(t,\tbar)
= {1\over (i\mu)^n} (\det\, A)^{-i\mu}\, {\rm
tr}\left({\partial\over\partial A}\right)^n (\det\,
A)^{i\mu}\, Z_{W_\infty}(t,\tbar)
\label{rewrit}
\end{equation}
It is easy to check explicitly that this constraint is solved by the
random-matrix model with partition function
\begin{equation}
Z_(t,\tbar) = (\det A)^\nu \,\int dM~ {\rm e}^{-\nu
\tr M A + (\nu - N) \tr\log M -\nu  \sum_{k>0}\, \tbar_k
\tr M^k}
\label{solv}
\end{equation}
where $\nu=-i\mu$. Differentiating in $\tbar_n$ and setting
$\tbar=0$, and dividing by the partition function, we find the
normalized correlator
\begin{equation}
\nu\langle \tr M^n \rangle \equiv
{1\over Z(A)} \int dM~\nu\tr M^n\, {\rm e}^{-\nu
\tr M A + (\nu - N) \tr\log M}
\label{norm}
\end{equation}
where now $Z$ is the partition function of the matrix model defined in
Eq.(\ref{finalmod}) above. This shows that the expectation values of
$\tr M^n$ computed via the quantum Penner action are (after
multiplying by $\nu$) just the expectation values of tachyons of
negative momentum, $\langle T_{-n}\rangle$, in the $c=1$ string at
self-dual radius. The same can be done for higher derivatives in
$\tbar$, so we have proved that the simple matrix model defined in the
previous section is (at least perturbatively) equivalent to this
string theory.

\section{Discussion and Conclusions}

A more detailed discussion of the model above, and in particular its
relation to previous attempts at finding a topological matrix model
for the $c=1$ string, can be found in Ref.\cite{im}. Here we point out
some of the interesting open questions that follow from this work.

Although the equivalence of our matrix model to the $c=1$ string has
been demonstrated only perturbatively, the matrix model seems
perfectly well-defined outside perturbation theory (it is really no
more than the matrix analogue of the Gamma-function). Thus one may ask
whether this provides new insight into the tricky question of giving a
nonperturbative definition to the $c=1$ string.

Recently Ghoshal and Vafa\cite{gv} have argued that the $c=1$ string
at self-dual radius describes universal properties of type II
superstrings compactified on Calabi-Yau manifolds which are developing
conifold singularities. Their identification makes use only of the
$c=1$ partition function. It would be interesting to understand
whether the amplitudes of the $c=1$ string likewise tell us something
about conifold singularities and related issues concerning
Ramond-Ramond states. It might, for example, be possible to address
some recent speculations due to Shenker\cite{shenker} using our
topological matrix model.

Finally, the simple way in which the background matrix $A$ enters into
our model suggests that one might understand something more about
background-independent string field theory from this point of view. In
the absence of a background of tachyons, the potential for our matrix
model is 0, rather analogous to the situation in topological field
theories, which are believed to contain background-independent
information about quantum gravity. It has been argued by us in
Ref.\cite{im} that this model contains other ``vacua'' corresponding
to string theory in the background of $c<1$ minimal models, so that
one could at least hope to find a background-independent picture for
all $c\le1$ string backgrounds.

\section{Acknowledgements}

One of us (S.M.) is grateful to the organisers of the Cargese School
for their kind invitation to lecture on this work, and for their
generous hospitality at Cargese.

\end{document}